\documentclass[%
superscriptaddress,
 amsmath,amssymb,
 aps,
 prd,
floatfix,nofootinbib
]{revtex4-1}

\usepackage{graphicx}
\usepackage{dcolumn}
\usepackage{bm}
\usepackage[normalem]{ulem}
\usepackage{tikz}
\usetikzlibrary{positioning}

\usepackage{braket}
\usepackage{multirow}
\usepackage{dsfont}
\usepackage{mathtools}
\usepackage{xcolor}
\usepackage[normalem]{ulem}

\usepackage{cancel}
\usepackage{epsfig,dsfont}
\usepackage{subfigure}
\usepackage{bm}
\usepackage{amssymb}
\usepackage{mathrsfs}
\usepackage{amsmath}
\usepackage{enumitem}
\usepackage{dcolumn}

\begin{document}

\preprint{KEK-CP-360}
\preprint{RBRC 1243}

\title{Symmetries of spatial meson correlators in high temperature QCD}

\author{\vspace*{4mm}C.~Rohrhofer}
\affiliation{Institute of Physics, University of Graz, 8010 Graz, Austria}
\author{Y.~Aoki}
\affiliation{KEK Theory Center, High Energy Accelerator Research Organization (KEK), Tsukuba 305-0801, Japan}
\affiliation{RIKEN BNL Research Center, Brookhaven National Laboratory, Upton NY 11973, USA}
\author{G.~Cossu}
\affiliation{School of Physics and Astronomy, The University of Edinburgh, Edinburgh EH9 3JZ, United Kingdom}
\author{H.~Fukaya}
\affiliation{Department of Physics, Osaka University, Toyonaka 560-0043, Japan}
\author{C.~Gattringer}
\affiliation{Institute of Physics, University of Graz, 8010 Graz, Austria}
\author{L.Ya.~Glozman}
\affiliation{Institute of Physics, University of Graz, 8010 Graz, Austria}
\author{S.~Hashimoto}
\affiliation{KEK Theory Center, High Energy Accelerator Research Organization (KEK), Tsukuba 305-0801, Japan}
\affiliation{School of High Energy Accelerator Science, The Graduate University for Advanced Studies (Sokendai), Tsukuba 305-0801, Japan}
\author{C.B.~Lang}
\affiliation{Institute of Physics, University of Graz, 8010 Graz, Austria}
\author{S.~Prelovsek\vspace{4mm}}
\affiliation{Faculty of Mathematics and Physics, University of Ljubljana, 1000 Ljubljana, Slovenia}
\affiliation{Jozef Stefan Institute, 1000 Ljubljana, Slovenia}
\affiliation{Institute f\"ur Theoretische Physik, Universit\"at Regensburg, D-93040, Germany\vspace*{4mm}}
 
\date{\today}

\begin{abstract}
\vspace{4mm}
Based on a complete set of $J = 0$ and $J=1$ spatial isovector correlation
functions calculated with $N_F = 2$ domain wall fermions  we identify an
intermediate temperature regime  of $T \sim 220 - 500$ MeV
($1.2T_c$--$2.8T_c$),
where chiral symmetry is restored but the 
correlators are not yet compatible with a simple free quark behavior.
More specifically, in the temperature range $T \sim 220 - 500$ MeV 
we identify a multiplet structure of spatial correlators that suggests
emergent $SU(2)_{CS}$ and $SU(4)$ symmetries, which are not symmetries of the
free Dirac action. The symmetry breaking effects in this temperature
range are less than 5\%. Our results indicate that  at these temperatures the chromo-magnetic 
interaction is suppressed and the elementary degrees of freedom are chirally symmetric quarks 
bound into color-singlet objects by the chromo-electric component of the gluon field.  At  temperatures between  500 and 660 MeV 
the emergent $SU(2)_{CS}$ and $SU(4)$ symmetries disappear  and one observes
a smooth transition to the regime
above $T \sim 1$ GeV where only chiral symmetries survive, which are finally compatible with 
quasi-free quarks.
\end{abstract}
\maketitle


\section{\label{sec:intro}Introduction}

Understanding the physics of strongly coupled matter at high temperature is one of the great open challenges in high
energy physics. Addressing this question is the subject of large-scale  experimental and theoretical efforts.
Initially it was assumed that above some pseudo-critical temperature $T_c$ quarks deconfine and chiral symmetry 
is restored such that above $T_c$ the degrees of freedom  are liberated quarks and gluons \cite{Sh}. 

A flavor non-singlet chiral restoration was indeed confirmed on the lattice, which is signalled by 
the vanishing quark condensate above the cross-over region around $T_c$
and by degeneracy of correlators that are connected by chiral transformations.  

The expected confinement-deconfinement transition turned out to be more intricate
to define. Such a transition was historically assumed to be associated with a different
expectation value of the Polyakov loop \cite{P,L} below and above the critical
temperature $T_c$. In pure $SU(3)$ gauge theory the Polyakov loop is connected
with the $Z_3$ center symmetry and indeed a sharp first-order phase transition
is observed \cite{Kaczmarek:2002mc}, which indicates that the relevant degrees of freedom below and above
$T_c$ are different. Still, one may ask whether this $Z_3$ transition is really connected with deconfinement in 
a pure glue theory. Traditionally the answer was affirmative, because the expectation value of the 
Polyakov loop can be related to the free energy of a static quark source. If this
energy is infinite, which corresponds to a vanishing Polyakov loop,
then we are in a confining mode, while deconfinement should be
associated with a finite free energy, i.e., a non-zero Polyakov loop.
However, this argumentation is  self-contradictory because a criterion for deconfinement in pure
gauge theory, i.e., deconfinement of gluons, is reduced to deconfinement
of a static charge (heavy quark), that is not part of the pure glue theory. 
The Polyakov loop is a valid order parameter but strictly speaking its relation to confinement is an assumption. And
indeed, just above the first-order $Z_3$  phase transition the energy and
pressure are quite different from the Stefan-Boltzmann limit which is associated
with free deconfined gluons \cite{b}.

In a theory with dynamical quarks the first-order phase transition is
washed out and on the lattice one observes a very smooth increase
of the Polyakov loop \cite{Petreczky:2015yta}. The reason for that behavior is rather clear:
in a theory with dynamical quarks there is no $Z_3$ symmetry and the
Polyakov loop ceases to be an order parameter.
Considering the finite energy of a pair of static quark sources (Polyakov loop correlator)
the resulting string breaking potential is due to vacuum loops of light quarks that combine 
with the static sources to a pair of heavy-light mesons.
Lattice measurements of the energy density and pressure with dynamical quarks indicate a smooth transition, and at $T \sim 1$ GeV the system is still quite far from the Stefan-Boltzmann
limit \cite{Karsch:2000ps,Bazavov:2017dsy}.

In view of the absence of a reliable, generally accepted definition and order parameter for deconfinement 
-- except for the most straightforward statement that confinement is the absence of colored states in the spectrum --
a key to understanding the nature of hot QCD matter is information
about the relevant effective degrees of freedom in high temperature QCD.
Several model and lattice studies suggest the
possible existence of inter-quark correlations or bound states
above $T_c$, see, e.g., Refs. \cite{Shuryak:2003ty,Ratti:2011au,Mukherjee:2015mxc}. 
While models may provide helpful intuitive understanding, it is important
to attempt finding model independent ways to identify
the degrees of freedom in high T QCD.

Among other observables, relevant information is encoded in Euclidean correlation
functions. 
At zero temperature hadron masses can be extracted from the exponential slope
of correlators in the Euclidean time direction $t$. At non-zero temperature
the temporal extent is finite by definition
(it vanishes at $T \rightarrow \infty$)
such that there is no strict notion of an asymptotic behavior for
$t$-correlators. 
Spatial correlators on the other hand are well-defined and do provide detailed
information about the QCD dynamics 
\cite{DeTar:1987xb,Born:1991zz,Fl,Kogut:1998rh,Pushkina:2004wa,Wi,Gavai:2006fs,Cheng:2010fe,Banerjee:2011yd}.
These spatial correlators can be analyzed  with respect to the symmetries they
exhibit, which in turn allows one to extract information about the relevant
effective degrees of freedom. 

In previous work \cite{Rohrhofer:2017grg} we have studied a complete
set of $J=0$ and $J=1$ isovector correlation functions in $z$-direction for a system 
with $N_F=2$ dynamical quarks in simulations with the chirally symmetric domain wall
Dirac operator at temperatures up to $T\sim 380$ MeV. Similar ensembles have been used previously for the
study of the $U(1)_A$ restoration in $t$-correlators and via the Dirac eigenvalue decomposition of correlators 
\cite{Cossu:2015kfa,Tomiya:2016jwr}. We have observed the restoration of both
$SU(2)_L \times SU(2)_R$ and $U(1)_A$ chiral symmetries at $T_c$ on a finite lattice of a given size.

However, by analyzing the formation of multiplets for the spatial correlators
even larger symmetries, referred to as $SU(2)_{CS}$
chiral spin and $SU(4)$ symmetries \cite{G1,GP}, have been identified in the 
$J=1$ correlators in the region $T \sim 2\,T_c$. These symmetries, while not
symmetries of the Dirac Lagrangian, are symmetries of the Lorentz-invariant
fermion charge. In the given reference frame they are symmetries of the interaction between
the chromo-electric field with the quarks while the interaction of quarks with the chromo-magnetic field breaks them.
These symmetries include as subgroups the chiral symmetries as well as
rotations between the right- and left-handed components of quarks.
Such symmetries have been found already earlier in the hadron spectrum at zero temperature \cite{D1,D2,D3,D4}
upon artificial truncation of the near-zero modes of the Dirac operator \cite{LS}. 
While the  $SU(2)_L \times SU(2)_R$ and $U(1)_A$ chiral symmetries are 
almost exact above $T_c$, the $SU(2)_{CS}$ and $SU(4)$ symmetries are approximate. In this paper we improve the analysis and extend the temperature range up to $T \sim 1$~GeV,
in order to further study the temperature evolution of the symmetries of correlators and thus the temperature
evolution of the emergent effective degrees of freedom.

We stress that the $SU(2)_{CS}$ and $SU(4)$ symmetries are not symmetries of
the free Dirac action and therefore their emergence is incompatible with the
notion of quasi-free, deconfined quarks. 
The emergence of these symmetries in a range from $T \sim 220$ -- $500$ MeV ($1.2T_c$ -- $2.8T_c$), as reported in this article,
suggests that the effective degrees of freedom of QCD at these temperatures are quarks with definite 
chirality bound by the chromo-electric component of the gluon field into color-singlet objects,
``string-like'' compounds. 

While the lattice study is possible only at zero chemical potential,
the observed approximate symmetries should persist also at finite chemical
potential, due to the quark chemical potential term in the QCD action
being manifestly $SU(2)_{CS}$ and $SU(4)$ symmetric \cite{G2}.

When increasing the temperature to
$T \sim 1$ GeV we observe that at very high temperature the
$SU(2)_{CS}$ and $SU(4)$ multiplet structure is washed out and the
full QCD meson correlators approach the corresponding correlators constructed with free,
non-interacting quarks. This indicates that at very high temperature
the coupling constant is sufficiently small to describe dynamics of weakly interacting quarks and gluons. Preliminary results of
this work were presented at the Lattice 2018 conference 
\cite{Rohrhofer:2018pey}.

\section{Spatial finite temperature meson correlators for non-interacting quarks in the continuum}
\label{sec:free}

We begin our presentation with a summary of the calculation of the spatial correlators for free massless 
quarks in the continuum. This situation is the limiting case that should represent QCD at very high temperatures 
where, due to asymptotic freedom, the interaction via gluons can be neglected. We discuss the multiplet 
structure for this reference case which we will later use to compare to our lattice calculation at high, but not 
asymptotically high temperature. In particular we will find that at moderately high temperatures above $T_c$ 
the spatial correlators of full QCD display a multiplet structure different from the limiting case of free quarks
discussed in this section. We remark that some of the free spatial continuum correlators computed here were 
already presented in \cite{Fl,Wi}, but for a systematical and complete discussion we need the full set of all spatial 
meson correlators and thus briefly summarize their derivation in this section and the appendix.

In the continuum the free spatial meson correlators in infinite spatial volume are given by 
\begin{equation}
C_\Gamma(z) \; = \; \int_{-\infty}^\infty \!\!dx  \int_{-\infty}^\infty \!\!dy  \int_{0}^\beta \!\!dt \; 
\left \langle {\cal O}_\Gamma(x,y,z,t) \; {\cal O}_\Gamma(0,0,0,0)^\dagger \right \rangle \; .
\label{corrdef}
\end{equation}
We consider Euclidean space at finite temperature, i.e., $x,y,z \in \mathds{R}$, and $t \in [0,\beta)$, where 
$\beta$ is the inverse temperature. In the correlators (\ref{corrdef}) we look at correlation in one of the spatial 
directions, here chosen as $z$, while the other two, $x$ and $y$, as well as the Euclidean time $t$ are integrated 
over. The latter integration over all coordinates that are perpendicular to the direction of propagation, i.e.,
the $z$-direction, fixes a ``Euclidean rest frame'' for our correlators.

The meson interpolators are given by 
\begin{equation}
{\cal O}_\Gamma({\bf x}) \; \equiv \; \overline{u} ({\bf x}) \Gamma d({\bf x}) \; , \quad
{\cal O}_\Gamma({\bf 0})^\dagger \; \equiv \; - \; \overline{d} ({\bf 0}) \Gamma^\dagger u({\bf 0}) 
 \; ,
\end{equation}
where we use the abbreviations ${\bf x} = (x,y,z,t)$ and  ${\bf 0} = (0,0,0,0)$, and $\Gamma$ is an element of
the Clifford algebra, i.e., a product of $\gamma$ matrices (see below). Note that choosing the negative sign for
${\cal O}_\Gamma^\dagger$ is a definition, since in general the sign obtained from conjugation will depend on $\Gamma$.
Throughout the whole paper we use the set $\gamma_\mu, \mu = 1,2,3,4$
of Euclidean $\gamma$-matrices that satisfy the anti-commutation relations
\begin{equation}
\gamma_\mu \gamma_\nu + \gamma_\nu \gamma_\mu \; = \; 
2\delta_{\mu \nu}\; , \; \qquad \gamma_5 \; \equiv \; \gamma_1\gamma_2\gamma_3\gamma_4 \; .
\label{eq:diracalgebra}
\end{equation}
 $\overline{u}({\bf x})$, $u({\bf x})$, $\overline{d}({\bf x})$, $d({\bf x})$ are free massless Dirac spinors 
which obey anti-periodic boundary conditions in Euclidean time. 
We remark that for simplicity we here have already expressed the non-singlet correlators in terms of the flavor 
spinors $u$ and $d$, while in the next section we write them in terms of isospin doublets 
$q({\bf x}) \equiv (u({\bf x}),d({\bf x}))$. 
After contracting the fermions, the two forms for writing the non-singlet bilinears of course give the same expressions. 

Performing these contractions we obtain 
\begin{equation}
\left \langle {\cal O}_\Gamma({\bf x}) \; {\cal O}_\Gamma({\bf 0})^\dagger \right \rangle \; = \; 
\mbox{Tr} \, \big[ S({\bf x},{\bf 0}) \, \Gamma^\dagger S({\bf 0},{\bf x}) \, \Gamma \big] \; ,
\label{contraction}
\end{equation}
where the trace is over Dirac indices and $S$ denotes the  
free continuum Dirac propagator. We are interested in the physics near
the chiral limit, and therefore we consider massless quarks in this section. In terms of
Fourier integrals $S$ is given by
\begin{equation}
S({\bf x},{\bf x}^\prime) \; = \; \frac{1}{(2\pi)^3 \beta} \int_{-\infty}^\infty \!\!\!\!dp_x  \int_{-\infty}^\infty \!\!\!\!dp_y 
\int_{-\infty}^\infty \!\!\!\!dp_z  
\sum_{n \in \mathds{Z}}  \, i \, \frac{\cancel{\bf p}}{{\bf p}^2} \, e^{\, i \,  {\bf p} \, ( {\bf x} -{\bf x}^\prime)} \; ,
\label{quarkprop}
\end{equation}
where ${\bf p} = (p_x,p_y,p_z,\omega_n)$, with the Matsubara frequencies 
$\omega_n = \pi(2n + 1)/\beta$. Inserting (\ref{quarkprop}) into 
(\ref{contraction}) and this into (\ref{corrdef}) we find
\begin{equation}
C_\Gamma(z) \; = \; -\frac{1}{(2\pi)^4 \beta} 
\int_{-\infty}^\infty \!\!\!\!dp_x  \int_{-\infty}^\infty \!\!\!\!dp_y \sum_{n \in \mathds{Z}} 
\int_{-\infty}^\infty \!\!\!\!dp_z \, \frac{e^{izp_z}}{{\bf p}^2} \int_{-\infty}^\infty \!\!\!\!dp_z^\prime \, 
\frac{e^{-izp_z^\prime}}{\widetilde{\bf p}^{\,2}} \; 
\mbox{Tr} \big[ \, \cancel{\bf p} \, \Gamma^\dagger \, \widetilde{\cancel{ \bf p}} \,  \Gamma \, \big] ,
\end{equation}
where $\widetilde{\bf p} \equiv (p_x,p_y,p_z^\prime,\omega_n)$ and we have already integrated over $x$, $y$ and 
$t$ in (\ref{corrdef}) which generated two Dirac deltas and a Kronecker delta that were used to get 
rid of two of the momentum integrals and one of the Matsubara sums. 

As we will see below, the trace in the integrand has the general form 
\begin{equation}
\mbox{Tr} \big[ \, \cancel{\bf p} \, \Gamma^\dagger \, \widetilde{\cancel{\bf p}} \,  \Gamma \, \big] \; = \; 
4 \big [ \, s_x \, p_x^2 \, + \, s_y \, p_y^2 \, + \,  s_z \, p_z \, p_z^\prime \, + \, s_\tau \, \omega_n^2 \, \big] \; ,
\label{tracewithsigns}
\end{equation}
where $s_x, s_y, s_z$ and $s_\tau$ are signs that depend on the choice of $\Gamma$. Thus for the pair of integrals
over the $z$ components we can distinguish two cases, depending on whether the factor $p_z \, p_z^\prime$ appears
in the integrand or not,
\begin{eqnarray}
\hspace*{-8mm} && \int_{-\infty}^\infty \!\!\!\!dp_z \, \frac{e^{izp_z}}{p_z^2 + \Omega^2} 
\int_{-\infty}^\infty \!\!\!\!dp_z^\prime \, \frac{e^{-izp_z^\prime}}{{p_z^{\prime}}^2 + \Omega^2} \; = \;  
\left[ \int_{-\infty}^\infty \!\!\!\!dp_z \, \frac{e^{izp_z}}{p_z^2 + \Omega^2} \right]^2  \; \equiv \; I_0^{\,2} \; , 
\\
\hspace*{-8mm} && \int_{-\infty}^\infty \!\!\!\!dp_z \, \frac{ e^{izp_z} \, p_z}{p_z^2 + \Omega^2} 
\int_{-\infty}^\infty \!\!\!\!dp_z^\prime \, \frac{e^{-izp_z^\prime} \, p_z^\prime}{{p_z^{\prime}}^2 + \Omega^2} \; = \;  
- \left[ \int_{-\infty}^\infty \!\!\!\!dp_z \, \frac{e^{izp_z} \, p_z}{p_z^2 + \Omega^2} \right]^2  \; \equiv \; - I_1^{\,2} \; ,
\end{eqnarray}
where we have defined $\Omega = \sqrt{ p_x^2 + p_y^2 + \omega_n^2}$. The integrals $I_0$ and $I_1$ are 
straightforward to solve with the residue theorem, 
\begin{equation}
I_0^{\,2} \; = \; \frac{\pi^2}{\Omega^2} \, e^{\, - z \, 2 \Omega} \; , \quad 
- I_1^{\,2} \; = \; \pi^2 \, e^{\, - z \, 2 \Omega} \; .
\end{equation}
We find for the correlator $C_\Gamma(z)$,
\begin{equation}
C_\Gamma(z) \; = \; - \big[ \, s_x \, C_s(z) \, + \, s_y \, C_s(z) \, + \,  s_z \, C_z(z) \, + \, s_\tau \, C_\tau(z) \, \big] \; , 
\label{C_Gamma_composition}
\end{equation}
with the individual correlators given by
\begin{eqnarray}
C_s(z) & = & \frac{1}{(2\pi)^2 \beta} \sum_{n \in \mathds{Z}} \int_{-\infty}^\infty \!\!\!\!dp_x  \int_{-\infty}^\infty \!\!\!\!dp_y \; 
\frac{e^{\, - 2\, z  \sqrt{ p_x^2 + p_y^2 + \omega_n^2}}}{p_x^2 + p_y^2 + \omega_n^2} \, p_x^2 \; ,
\label{CsCzCtau}  
  \\
C_z(z) & = & \frac{1}{(2\pi)^2 \beta}  \sum_{n \in \mathds{Z}} \int_{-\infty}^\infty \!\!\!\!dp_x  \int_{-\infty}^\infty \!\!\!\!dp_y \; 
e^{\, - 2\,z  \sqrt{ p_x^2 + p_y^2 + \omega_n^2}}  \; ,
\nonumber \\
C_\tau(z) & = & \frac{1}{(2\pi)^2 \beta}  
 \sum_{n \in \mathds{Z}} \int_{-\infty}^\infty \!\!\!\!dp_x  \int_{-\infty}^\infty \!\!\!\!dp_y \; 
\frac{e^{\, - 2\,z \sqrt{ p_x^2 + p_y^2 + \omega_n^2}}}{p_x^2 + p_y^2 + \omega_n^2} \, \omega_n^{\,2}  \; .
\nonumber 
\end{eqnarray}
The correlators $C_s(z), C_z(z)$ and $C_\tau(z)$ obey the obvious sum rule
\begin{equation}
2 \, C_s(z) \; + \; C_\tau(z) \; = \; C_z(z) \; ,
\label{sumrule}
\end{equation}
i.e., only two of them are independent. We choose $C_z(z)$ and $C_\tau(z)$ to express all other correlators.
The treatment of the Matsubara sums and the necessary integrals for evaluating $C_z(z)$ and $C_\tau(z)$
are discussed in Appendix A, where we also discuss the asymptotic behavior of the correlators.

We now come to the identification of multiplets, i.e., we identify the sets of Clifford algebra elements
$\Gamma$ that share the same decay properties for their corresponding correlators $C_\Gamma(z)$. 
For this we need to determine the signs $s_x, s_y, s_z$ and $s_\tau$ in the traces 
(\ref{tracewithsigns}) for the different choices of $\Gamma$, which in turn determine how the 
respective correlator $C_\Gamma(z)$ is composed from the contributions $C_s(z)$, $C_z(z)$ and $C_\tau(z)$ 
according to (\ref{C_Gamma_composition}).

We first note that for chiral partners, i.e., correlators where $\Gamma$ is replaced by $\Gamma \gamma_5$,
the corresponding correlators $C_\Gamma(z)$ and $C_{\Gamma \gamma_5}(z)$ have opposite overall signs, 
and thus also opposite individual signs $s_x, s_y, s_z$ and $s_\tau$. This follows from the trivial relation
\begin{equation}
\mbox{Tr} \big[ \, \cancel{\bf p} \, (\Gamma \gamma_5)^\dagger \, \widetilde{\cancel{\bf p}} \,  \Gamma \gamma_5 \, \big]
\; = \; - \; \mbox{Tr} \big[ \, \cancel{\bf p} \, \Gamma^\dagger \, \widetilde{\cancel{\bf p}} \,  \Gamma \, \big] \; .
\label{g5signflip}
\end{equation}
This implies that we need to determine the signs $s_x, s_y, s_z$ and $s_\tau$ in the traces 
(\ref{tracewithsigns}) only for 8 out of the 16 Clifford algebra generators $\Gamma$. 
Our results for the signs $s_x, s_y, s_z$ and $s_\tau$ that determine the decomposition of 
$\mbox{Tr} \big[ \, \cancel{\bf p} \, \Gamma^\dagger \, \widetilde{\cancel{\bf p}} \,  \Gamma \, \big]$ according to 
(\ref{tracewithsigns}) are listed in Table~\ref{table:gammasigns}. 

\begin{table}[t]
\begin{center}
\begin{tabular}[t]{c|cccc|cc}
\qquad $\;\,\Gamma$ \qquad \quad&   
\quad $\!\!\!s_x$  \quad &  \quad $\!\!\!s_y$  \quad &  \quad $\!\!\!s_z$  \quad &  \quad $\!\!\!s_\tau$  \quad 
& \qquad name \qquad  & \quad chiral partner \quad \\[1ex]
\hline
\rule{0pt}{3ex}
$\mathds{1}$ & $+$ & $+$ & $+$ & $+$ & \quad $S$ & $PS$ \\[1ex]
\hline
\rule{0pt}{3ex}
$\!\gamma_1$ & $+$ & $-$ & $-$ & $-$ & \quad $V_x$ & $A_x$ \\
$\gamma_2$ & $-$ & $+$ & $-$ & $-$ & \quad $V_y$ & $A_y$ \\
$\gamma_4$ & $-$ & $-$ & $-$ & $+$ & \quad $V_t$ & $A_t$ \\[1ex]
\hline        
\rule{0pt}{3ex}
$\!\gamma_1 \gamma_3$ & $-$ & $+$ & $-$ & $+$ & \quad $T_x$ & $X_x$ \\
$\gamma_2 \gamma_3$ & $+$ & $-$ & $-$ & $+$ & \quad $T_y$ & $X_y$ \\
$\gamma_4 \gamma_3$ & $+$ & $+$ & $-$ & $-$ & \quad $T_t$ & $X_t$ \\[1ex]
\hline
\rule{0pt}{3ex}
$\gamma_3$ & $-$ & $-$ & $+$ & $-$ &  &  \\
\end{tabular}
\end{center}
\caption{The signs $s_x$, $s_y$ $s_z$ and $s_\tau$ that determine the trace  
$\mbox{Tr} \big[ \, \cancel{\bf p} \, \Gamma^\dagger \, \widetilde{\cancel{ \bf p}} \,  \Gamma \, \big]$
for different choices of $\Gamma$ according to (\ref{tracewithsigns}). For chiral partners, i.e., when 
$\Gamma$ is replaced by $\Gamma \gamma_5$, all signs are reversed (compare (\ref{g5signflip})). 
To simplify the notation we chose the (irrelevant) overall signs equal for both chiral partners such that 
the relative signs $s_x$, $s_y$ $s_z$ and $s_\tau$ as listed in the table are used for both chiral partners.
In the two columns on the right we give the names of the bilinears and their chiral partners
which we will discuss in detail in the next section. Since the interpolators with $\gamma_3$ and 
$\gamma_3 \gamma_5$ vanish identically no name is assigned.
\label{table:gammasigns}}
\end{table}

Having determined the signs $s_x, s_y, s_z, s_\tau$ we use them in (\ref{C_Gamma_composition}) to work out the 
composition of $C_\Gamma(z)$ from the building blocks $C_s(z)$, $C_z(z)$ and $C_\tau(z)$, 
and after  eliminating $C_s(z)$ we obtain the representation for the  $C_\Gamma(z)$ in terms of 
$C_z(z)$ and $C_\tau(z)$ evaluated in  Appendix A. We find (overall signs were chosen such that chiral partners 
have the same overall sign),
\begin{eqnarray}
\hspace*{-6mm} &&   \; \; C_{\mathds{1}}(z)  =  C_{\gamma_5}(z) \; = \; 
2 C_s(z) \, + \, C_z(z) \, + \, C_\tau(z) \; = \; 2 \, C_z(z) \; , 
\label{Corr_multiplets}
\\
\hspace*{-6mm} &&   \; \; C_{\gamma_1}(z) = C_{\gamma_1 \gamma_5}(z) =  
C_{\gamma_2}(z) = C_{\gamma_2 \gamma_5}(z) \; = \; C_z(z) + C_\tau(z)  \; , 
\nonumber \\
\hspace*{-6mm} &&   \; \; C_{\gamma_4}(z) = C_{\gamma_4 \gamma_5}(z) 
=  \; 2 C_s(z) + C_z(z) - C_\tau(z) \; = \; 2(C_z(z) -  C_\tau(z)) \; , 
\nonumber \\
\hspace*{-6mm} &&  \; \; C_{\gamma_1\gamma_3}(z) = C_{\gamma_1 \gamma_3 \gamma_5}(z) =  
C_{\gamma_2 \gamma_3 }(z) = C_{\gamma_2 \gamma_3 \gamma_5}(z) \; = \; C_z(z) - C_\tau(z) \; , 
\nonumber \\
\hspace*{-6mm} &&  \; \; C_{\gamma_4 \gamma_3}(z) = C_{\gamma_4 \gamma_3\gamma_5}(z) =  
\; - 2 C_s(z) + C_z(z) + C_\tau(z) \; = \; 2\,  C_\tau(z) \; , 
\nonumber \\
\hspace*{-6mm} &&  \; \; C_{\gamma_3}(z)  =  C_{\gamma_3 \gamma_5}(z) \; = \; 
2 C_s(z) \, - \, C_z(z) \, + \, C_\tau(z) \; = \; 0 \; . 
\nonumber
\end{eqnarray}
The vanishing of the correlators $C_{\gamma_3}(z)$ and $C_{\gamma_3 \gamma_5}(z)$ is a direct consequence 
of the sum rule (\ref{sumrule}). From a more physical point of view this vanishing is a consequence of current 
conservation. Indeed  $C_{\gamma_3}(z)$ is the correlator for the 3-component of the conserved vector current 
$J_\mu({\bf x}) = \overline{u}({\bf x}) \, \gamma_\mu d({\bf x})$ and concerning the propagation in $z$-direction
the integral $\int \!\! dx dy dt \, J_3(x,y,z,t)$ is a conserved charge. Thus the corresponding spatial correlator and 
its chiral partner vanish, which also implies that the sum rule (\ref{sumrule}) is directly linked 
to current conservation. Furthermore the sum rule (current conservation) means that the correlators
$C_{\gamma_4}(z) = C_{\gamma_4\gamma_5}(z)$
are not independent from the correlators $C_{\gamma_1\gamma_3}(z) = C_{\gamma_1 \gamma_3 \gamma_5}(z) =  
C_{\gamma_2 \gamma_3 }(z) = C_{\gamma_2 \gamma_3 \gamma_5}(z)$.
			    
We conclude this section with quoting the asymptotic behavior of our correlators, which is obtained by 
using (\ref{J_asymptotics}) from Appendix A in the expressions (\ref{Corr_multiplets}),
\begin{eqnarray}
\hspace*{-5mm}&& C_{\mathds{1}}(z) = \frac{2\pi}{\beta^3} \frac{e^{\, - \, 2 \, z \, \omega_0}}{2 \, z \, \omega_0}
\left[ 1 + \frac{1}{2 \, z \,  \omega_0} \right] +  
O \! \left( \frac{e^{\, - \, 4 \, z \, \omega_0}}{z\omega_0} \right) ,
\nonumber \\
\hspace*{-5mm}&&
\nonumber \\
\hspace*{-5mm}&& C_{\gamma_1}(z) = C_{\gamma_2}(z) = \frac{2\pi}{\beta^3}
\frac{e^{\, - \, 2 \, z \, \omega_0}}{2 \,  z \, \omega_0}
\left[ 1 + \frac{1}{(2 \, z \, \omega_0)^2} + ... \right] + 
O \! \left( \frac{e^{\, - \, 4 \, z \, \omega_0}}{z\omega_0} \right)  ,
\nonumber \\
\hspace*{-5mm}&&
\nonumber \\
\hspace*{-5mm}&& C_{\gamma_4}(z) = 
\frac{4\pi}{\beta^3} \frac{e^{\, - \, 2 \, z \, \omega_0}}{(2 \,z \, \omega_0)^2}
\left[ 1 - \frac{1}{2 \, z \, \omega_0} + ... \right] +  
O \! \left( \frac{e^{\, - \, 4 \, z \, \omega_0}}{z\omega_0} \right) ,
\nonumber \\
\hspace*{-5mm}&&
\nonumber \\
\hspace*{-5mm}&& C_{\gamma_1 \gamma_3}(z) = C_{\gamma_2 \gamma_3}(z) = 
\frac{2\pi}{\beta^3} \frac{e^{\, - \, 2 \, z \, \omega_0}}{(2 \, z \, \omega_0)^2}
\left[ 1 - \frac{1}{2 \, z \, \omega_0} + ... \right] +  
O \! \left( \frac{e^{\, - \, 4 \, z \, \omega_0}}{z\omega_0} \right)\!,
\nonumber \\
\hspace*{-5mm}&&
\nonumber \\
\hspace*{-5mm}&& C_{\gamma_4 \gamma_3}(z) = \frac{2\pi}{\beta^3}
\frac{e^{\, - \, 2 \, z \, \omega_0}}{2 \, z \, \omega_0}
\left[ 1 - \frac{1}{2 \, z \, \omega_0} + ... \right] +  
O \! \left( \frac{e^{\, - \, 6 \, z \, \omega_0}}{z\omega_0} \right)  ,
\nonumber \\
\hspace*{-5mm}&&
\nonumber \\
\hspace*{-5mm}&& C_{\gamma_3}(z) = 0 \; .
\label{Corr_multiplets_asympt}
\end{eqnarray} 
Here we have only listed half of the correlators in each chiral multiplet without their chiral partners, which have identical correlators (up to an overall sign which we dropped). The fact that on the rhs.\ of (\ref{Corr_multiplets_asympt})
appears only the dimensionless combination $z\omega_0 = \pi \, z/\beta = \pi \, z T $ reflects the absence
of any physical scale in the conformal theory of massless non-interacting
quarks.

\section{\label{sec:ops}Fermionic bilinears and their symmetries}

Having summarized the explicit form of the spatial correlators for the free case, let us now come to the general  
(full QCD) discussion of the mesonic bilinears and their symmetries.
We are interested in the spatial correlators of the local isovector mesonic bilinears 
\begin{align}
\mathcal{O}_\Gamma(x) \;  = \; \bar q(x) \,  \Gamma \, \frac{\vec{\tau}}{2} \, q(x) \; ,
\label{equ:localoperator}
\end{align}
which we now write using the isospin doublets $q(x) \equiv (u(x),d(x))$. The isovector structure of the 
bilinears is determined by the isospin Pauli matrices $\tau_a$. Again $\Gamma$ may be any element of the 
Clifford algebra and the choice of $\Gamma$ determines the symmetry properties of the respective bilinear.

Two $J=0$ bilinears can be defined by the following choices for $\Gamma$:
\begin{align}
\Gamma = \left\{
\begin{array}{cccl}
\gamma_5 & \quad \dots \; & PS & \quad (Pseudoscalar) \; ,\\
\mathds{1} & \quad \dots \; & S & \quad (Scalar) \; .
\end{array}\right.
\end{align}
These two bilinears can be transformed into each other by global ${U}(1)_A$ rotations
\begin{align}
q(x) \rightarrow \exp\left(i\gamma_5\theta\right)q(x) \; .
\label{equ:u1atrafos}
\end{align}

For $J=1$ we consider bilinears with the following choices of $\Gamma$ that define the 
{\it Vector} bilinears $\mathbf{V}$:
\begin{align}
\Gamma = \left\{
\begin{array}{cccl}
\gamma_1 & \quad \dots \; & V_x \; , & \\
\gamma_2 & \quad \dots \; & V_y \; , & \quad \textrm{({\it Vector})} \\
\gamma_4 & \quad \dots \; & V_t \; . &
\end{array}\right.
\label{v}
\end{align}
As we have already seen for the free case which we discussed in the previous section, due to current conservation the 
3-component $\bar q(x) \gamma_3 \frac{\vec{\tau}}{2} q(x)$ does not propagate in the $z$ direction such that 
we omit the choice $\Gamma = \gamma_3$.

The vector bilinears are related to their chiral partners through flavor non-singlet axial rotations 
\begin{align}
q(x) \rightarrow \exp\left(\frac{i}{2}\gamma_5\vec\tau\vec\theta\right)q(x) \; .
\label{equ:su2trafos}
\end{align}
Their chiral partners, the {\it Axial-vector} bilinears $\mathbf{A}$ are defined as:
\begin{align}
\Gamma = \left\{
\begin{array}{cccl}
\gamma_1\gamma_5 & \quad \dots \; & A_x \; , & \\
\gamma_2\gamma_5 & \quad \dots \; & A_y \; , & \quad \textrm{({\it Axial-vector})} \\
\gamma_4\gamma_5 & \quad \dots \; & A_t \; . &
\end{array}\right.
\label{av}
\end{align}
At zero (or sufficiently small) temperature the chiral partner of the non-propagating third vector current component, 
i.e., the bilinear with the gamma structure $\Gamma=\gamma_3\gamma_5$,
does indeed propagate also in $z$-direction due to broken chiral
symmetry and then couples to the pseudoscalar channel. After restoration of chiral symmetry, i.e., 
at the temperatures we consider here, it behaves like its chiral partner and does not propagate in $z$-direction.
Thus, like $\Gamma = \gamma_3$, also the choice $\Gamma = \gamma_3\gamma_5$ can be omitted.

The bilinears that correspond to the six tensor elements $\sigma_{\mu\nu}$ of the Clifford algebra can be organized
into two vector-valued objects, the {\it Tensor-vector}~$\mathbf{T}$:
\begin{align}
\Gamma = \left\{
\begin{array}{cccl}
\gamma_1\gamma_3 & \quad \dots \; & T_x \; , & \\
\gamma_2\gamma_3 & \quad \dots \; & T_y \; , & \quad \textrm{({\it Tensor-vector})} \\
\gamma_4\gamma_3 & \quad \dots \; & T_t \; , &
\end{array}\right.
\label{t}
\end{align}
and the {\it Axial-tensor-vector}~$\mathbf{X}$:
\begin{align}
\Gamma = \left\{
\begin{array}{cccl}
\gamma_1\gamma_3\gamma_5 & \; \dots \; & X_x \; , & \\
\gamma_2\gamma_3\gamma_5 & \; \dots \; & X_y \; , & \quad \textrm{({\it Axial-tensor-vector})} \\
\gamma_4\gamma_3\gamma_5 & \; \dots \; & X_t \; . &
\end{array}\right.
\label{x}
\end{align}
The bilinears $\mathbf{T}$ and $\mathbf{X}$ can be transformed into each other by the $U(1)_A$ 
rotations (\ref{equ:u1atrafos}). Table~\ref{tab:ops} summarizes our bilinears and lists the $U(1)_A$ and 
$SU(2)_L \times SU(2)_R$ relations among them. 

Due to the restoration of the $U(1)_A$ and $SU(2)_L \times SU(2)_R$ symmetries at high temperature we 
expect the emergence of degeneracies among correlators of bilinears related by these symmetries, and of 
course those degeneracies clearly must also be seen explicitly in the free continuum correlators
(\ref{Corr_multiplets}), (\ref{Corr_multiplets_asympt}). The degeneracies based on $U(1)_A$ and 
$SU(2)_L \times SU(2)_R$ are the degeneracies required by chiral symmetries that emerge above $T_c$. 

\begin{table}
\center
\begin{tabular}{cccll}
\hline\hline
\rule{0pt}{3ex}
 Name        &
 Dirac structure &
 \quad Abbreviation  \quad &
 \multicolumn{2}{l}{
 } 
 
\\[1ex]
\hline
\rule{0pt}{3ex}

\textit{Pseudoscalar}        & $\gamma_5$                 & $PS$         &
 \multirow{2}{1cm}{$\left.\begin{aligned}\\ \end{aligned}\right] U(1)_A$}& \\
\textit{Scalar}              & $\mathds{1}$               & $S$        &  & 
\\[1ex]
\hline
\rule{0pt}{3ex}
\textit{Axial-vector}        & $\gamma_k\gamma_5$         & $\mathbf{A}$ & 
\multirow{2}{1cm}{$\left.\begin{aligned}\\ \end{aligned}\right] SU(2)_A$}&\\
\textit{Vector}              & $\gamma_k$                 & $\mathbf{V}$ & & \\
\textit{Tensor-vector}       & $\gamma_k\gamma_3$         & $\mathbf{T}$ & 
\multirow{2}{1cm}{$\left.\begin{aligned}\\ \end{aligned}\right] U(1)_A$} &\\
\textit{Axial-tensor-vector} & $\gamma_k\gamma_3\gamma_5$ & $\mathbf{X}$ & &\\[1ex]
\hline\hline
\end{tabular}
\caption{Fermion bilinears considered in this work and their transformation properties
(last column). This classification assumes propagation in $z$-direction. The
open vector index $k$ here runs over the components $1,2,4$, i.e., $x,y$ and $t$.}
\label{tab:ops}
\end{table}

However, in addition to those, at 
temperatures not too far above $T_c$ a larger group of symmetries, $SU(2)_{CS}$ and $SU(4)$ that contain 
$U(1)_A$ and $SU(2)_L \times SU(2)_R$ \cite{G1,GP},
\begin{equation}
SU(2)_{CS} \; \supset \; U(1)_A \qquad \mbox{and} \qquad  
SU(4)  \; \supset  \;  SU(2)_L \times SU(2)_R \times U(1)_A \; ,
\end{equation}
has been observed in our previous study of correlators \cite{Rohrhofer:2017grg}. 
The $SU(2)_{CS}$ chiral spin transformations are defined by
\begin{equation}
q(x) \; \rightarrow \; \exp\left(\frac{i}{2}\vec\Sigma \, \vec\epsilon\right)q(x) \; , \quad \;
\bar{q}(x) \; \rightarrow \; \bar{q}(x) \gamma_4 \exp\left(-\frac{i}{2}\vec\Sigma \, \vec\epsilon\right) \gamma_4 \; ,
\label{equ:su2cstrafos}
\end{equation}
where $\vec\epsilon \in \mathds{R}^3$ are the rotation parameters. For the generators $\vec\Sigma$ one has 
four different choices $\vec\Sigma = \vec\Sigma_k$ with $k = 1,2,3,4$, but, as we will discuss below, only the cases $k = 1$ and $k=2$ are of interest here. The generators are given by
\begin{align}
\vec\Sigma_k \; = \; \{\gamma_k,-i\gamma_5\gamma_k,\gamma_5\} \; ,
\end{align}
and the $su(2)$ algebra is  satisfied for any choice $k=1,2,3,4$.
While these are not symmetries of the Dirac lagrangian, both in
Minkowski and Euclidean space,
the Lorentz-invariant fermion charge in Minkowski space 
\begin{equation}
Q \;  = \; \int \!\! d^3x  \; 
\psi^\dagger(x)  \psi(x),
\label{Q}
\end{equation}
is invariant
under  $SU(2)_{CS}$,
where $\psi(x)$ can be either a single-flavor quark field or an isospin
doublet. The Euclidean fermion charge is also $SU(2)_{CS}$ invariant.

In Minkowski space in a given reference frame the quark-gluon interaction can be split into temporal and spatial parts:
\begin{equation}
\overline{\psi} \,  \gamma^{\mu} D_{\mu} \, \psi  \; = \; \overline{\psi}  \, \gamma^0 D_0  \, \psi 
\; + \; \overline{\psi} \,  \gamma^i D_i \, \psi \; ,
\label{cl}
\end{equation}
where 
\begin{equation}
D_{\mu}\psi \; = \; \left( \partial_\mu - ig \frac{{\bf t} \cdot {\bf A}_\mu}{2}\right )\psi \; .
\end{equation}
The temporal term includes the interaction of the color-octet charge density 
\begin{equation}
\bar \psi (x) \,  \gamma^0  \frac{{\bf t}}{2} \, \psi(x) \; = \; \psi (x)^\dagger \, \frac{{\bf t}}{2} \, \psi(x)
\label{den}
\end{equation}
with the chromo-electric component of the gluonic field. It is invariant  
under $SU(2)_{CS}$  \cite{GP}. We emphasize that the $SU(2)_{CS}$ transformations
defined in Eq.~(\ref{equ:su2cstrafos}) via the Euclidean
Dirac matrices can be identically applied to Minkowski Dirac spinors without
any modification of the generators.
The spatial part contains the quark kinetic term and the interaction with the chromo-magnetic field. This term 
breaks $SU(2)_{CS}$. In other words: the $SU(2)_{CS}$ symmetry distinguishes between quarks interacting 
with the chromo-electric and chromo-magnetic components of the gauge field. It is important to note that  
discussing ``electric'' and ``magnetic'' components can be done
only in Minkowski space and in addition one needs to fix the reference frame.  However, at high 
temperatures Lorentz invariance is broken and a natural frame to discuss physics is the rest frame of the medium. 

The $SU(2)_{CS}$ transformations (\ref{equ:su2cstrafos}) with $k=1$ generate the following two 
$SU(2)_{CS}$ - singlets and two $SU(2)_{CS}$ - triplets of bilinears:
\begin{align}
(V_y); \; (A_y, T_t, X_t) \; , \label{equ:e2a} \\
(V_t); \; (A_t, T_y, X_y) \; . \label{equ:e3a}
\end{align}
These irreducible representations of $SU(2)_{CS}$ can be obtained
by applying the $SU(2)_{CS}$ transformation (\ref{equ:su2cstrafos})
on any of the bilinears from the given representation and the result
will be a linear combination of all bilinears in the given representation.
The observation of a degeneracy of the correlators built from the triplet bilinears in Eq.~(\ref{equ:e2a}) would imply the
emergence of the corresponding $SU(2)_{CS}$ symmetry. We stress that this is not a symmetry of deconfined 
free quarks, see Eq.~(\ref{Corr_multiplets}), and the observation of a degeneracy within the triplet in Eq. (\ref{equ:e2a}) means that the quarks in the system interact exclusively via 
the chromo-electric field, without any chromo-magnetic admixture.
Since only color-singlet bilinears can propagate on the
lattice at any temperature the systems represent color-singlet 
quark - antiquark objects
bound by chromo-electric interactions.

Note that the observation of a degeneracy of correlators for the triplet bilinears in Eq.~(\ref{equ:e3a}) 
would not discriminate 
between the confining mode and free quarks, because the current conservation in the free quark system also 
provides such a degeneracy, as follows already from the discussion in the previous section, see 
Eq.~(\ref{Corr_multiplets})\footnote{This is true for the correlators normalized to 1 which we study here. Without this 
normalization there is an overall factor of $2$ between the free correlators built with the $V_t,A_t$ and 
$T_x,T_y,X_x,X_y$ bilinears (see, e.g., Eq.~(\ref{Corr_multiplets_asympt})), that would allow one to distinguish 
the results for free quarks from the full $SU(2)_{CS}$ case in an elaborated calculation with properly 
renormalized full QCD correlators.}.

The transformations (\ref{equ:su2cstrafos}) with $k=2$ generate the  following singlets and  triplets:
\begin{align}
 (V_x); \; (A_x, T_t, X_t) \; , \label{equ:e2b} \\
 (V_t); \;(A_t, T_x, X_x) \; . \label{equ:e3b}
\end{align}
Again, a degeneracy of the correlators built from the triplet bilinears in Eq.~(\ref{equ:e2b}) is a signal for the emergence 
of the $SU(2)_{CS}$ symmetry.  This is different from the degeneracy of the correlators of the triplet bilinears from 
Eq.~(\ref{equ:e3b}) which in the free quark case can be connected to current conservation and thus is not suitable for 
discriminating between the interacting mode and a system of free quarks.

This discussion (as well as a structure of the $SU(4)$ multiplets below) implies that only the study of a possible degeneracy among correlators of the bilinears (\ref{equ:e2a}), 
as well as the bilinears (\ref{equ:e2b}) is suitable for the analysis of the underlying dynamics and degrees of freedom.
Note that only those $SU(2)_{CS}, k=1,2,3,4$ transformations can be considered for a given observable 
that do not mix operators of different spin and thus respect rotational invariance at non-zero temperature. 
This requirement is met for our setup by the $k=1,2$ transformations, as indicated above.

We remark that at zero temperature in the continuum there is a $SO(3)$ symmetry in the $x,y,t$ 
subspace and the $z$-correlators of the $V_x,V_y,V_t$ bilinears (\ref{v}) coincide. The same is true for the 
$z$-correlators of the corresponding $x,y$ and $t$ components of the bilinears (\ref{av}), (\ref{t}) and (\ref{x}). 
At finite temperature this rotational symmetry is broken down to a residual $SO(2)$ symmetry which connects the 
correlators of the spatial components $V_x \leftrightarrow V_y$ and $A_x \leftrightarrow A_y$ et cetera. On the 
lattice the reduced symmetry for the $T>0$ case and the $z=const$ subspace is $D_{4h}$ and the relevant 
symmetry is $S_2 \times SU(2)_{CS}$ \cite{Rohrhofer:2017grg}\footnote{$S_2$ here denotes the 
\emph{permutation-} or \emph{symmetric group} for $x \leftrightarrow y$ interchanges.}, such that the multiplets are
\begin{align}
(V_x,V_y); \; (A_x, A_y, T_t, X_t) \; ,  \label{equ:s2a} \\
(V_t); \;   (A_t, T_x, T_y, X_x, X_y) \; . \label{equ:s2b} 
\end{align}

Finally we remark that the group $SU(2)_{CS} \otimes SU(2)_F$, where $SU(2)_F$ is the isospin symmetry
group, can be extended to $SU(4)$ with fifteen generators:
\begin{align}
\{ (\vec{\tau} \otimes \mathds{1}_D) ,
(\mathds{1}_F \otimes \vec \Sigma_k) ,
(\vec{\tau} \otimes \vec \Sigma_k) \} \; .
\label{gensSU4}
\end{align} 
The corresponding transformations are a trivial generalization of Eq.~(\ref{equ:su2cstrafos}) 
obtained by replacing the generators 
$\vec{\Sigma}$ by those listed in (\ref{gensSU4}).  Also the  group $SU(4)$ is a symmetry of the quark - chromo-electric
interaction terms of the QCD lagrangian, while the quark - chromo-magnetic
interaction as well as the kinetic term break it.
The $S_2 \times SU(4)$ transformations connect the following $J=1$
operators from Table \ref{tab:ops}:
\begin{align}
(V_x, V_y, A_x, A_y, T_t, X_t) \; , \label{equ:ss2a} \\
(V_t, A_t, T_x, T_y, X_x, X_y) \; . \label{equ:ss2b} 
\end{align}

These are the multiplets of the isovector operators that are discussed in the present paper.
The $SU(4)$ symmetry requires degeneracy within both, the
(\ref{equ:ss2a}) as well as the (\ref{equ:ss2b}) multiplets, while a degeneracy of the normalized correlators 
from the multiplet (\ref{equ:ss2b}) is also consistent with free non-interacting quarks.
Obviously the chiral multiplets of the $PS$ and $S$ bilinears are not subject to this degeneracy.

The complete $S_2 \times SU(4)$ multiplets in addition also include the isoscalar
partners of $A_x, A_y, T_t$ and $X_t$ in Eq.~(\ref{equ:ss2a}) as well as the isoscalar partners of 
$A_t, T_x, T_y, X_x$ and $X_y$ in Eq.~(\ref{equ:ss2b}). The isoscalar partners of $V_x, V_y$ and $V_t$ 
are the SU(4) singlets.


\section{\label{sec:sim}Lattice technicalities}

The correlators discussed in the previous section are evaluated on the JLQCD
configurations for full QCD with $N_F=2$ flavors of domain wall fermions.
Details concerning the gauge configurations are presented in \cite{Cossu:2015kfa,Tomiya:2016jwr}.
In this setup we choose $L_5$, the extent of the auxiliary 5-th dimension,
such that for all our ensembles the violation of the Ginsparg-Wilson condition
is less than $1$~MeV.

For measurements the IroIro software is used~\cite{Cossu:2013ola}, and the
relevant parameters are fixed in a zero temperature study~\cite{Kaneko:2013jla}.
The quark propagators are computed on point sources with the domain wall Dirac
operator after three steps of stout smearing. The fermion fields are periodic in
the spatial directions and anti-periodic in time.

We use the Symanzik-improved gauge action at inverse gauge couplings $\beta_g$
in a range between $\beta_g = 4.1$ and $\beta_g = 4.5$, and with the different
temporal lattice extents in use, $N_t = 4,6,8$ and $N_t = 12$, we cover a
range of temperatures between $T\simeq$ 220 MeV and $T\simeq$ 960 MeV.
For the bare quark mass parameters $m_u = m_d \equiv m_{ud}$ we use  the value
$m_{ud}=0.001$ which corresponds to physical quark masses at our different
temperatures in the range between 2~MeV and 4~MeV.
We have also performed simulations with $m_{ud}=0.01$, $m_{ud}=0.005$ and
observed stability of our results against quark mass variation because
in the temperature range we consider (220 -- 960 MeV) these quark masses are
essentially negligible due to temperature effects.
Further details concerning the chiral properties for our set of parameters are
given in~\cite{Cossu:2015kfa,Tomiya:2016jwr}.
The complete list of our ensembles and their parameters is provided in
Table~\ref{tab:ensembles}.

\begin{table}
\center
\begin{tabular}{cccccccc}
\hline\hline
\rule{0pt}{3ex}
\quad $\!\!N_s^3\times N_t$ \quad & \quad $\!\!\beta_g$ \quad & \quad $\!\!\!\!a$ [fm] \quad & \quad $\!\!\!\!m_{ud}$ \quad & \# configs \quad & \quad 
$\!\!\!L_5$ \quad & $T$ [MeV] & \quad $\!\!T/T_c$ \quad 
\\[1ex]
\hline
\rule{0pt}{3ex}
$\!\!32^3\times 12$ & $4.30$  & $0.075$& $0.001$ & 226  & 24  & $220$ & $1.2$ \\
$32^3\times 8$ & $4.10$  & $0.113$ & $0.001$ &     800  & 24  & $220$ & $1.2$ \\
$32^3\times 8$ & $4.18$  & $0.096$ & $0.001$ &     230  & 12  & $260$ & $1.5$ \\
$32^3\times 8$ & $4.30$  & $0.075$ & $0.001$ &     260  & 12  & $320$ & $1.8$ \\
$32^3\times 8$ & $4.37$  & $0.065$ & $0.001$ &      77  & 12  & $380$ & $2.2$ \\
$32^3\times 6$ & $4.30$  & $0.075$ & $0.001$ &     270  & 12  & $440$ & $2.5$ \\
$32^3\times 8$ & $4.50$  & $0.051$ & $0.001$ &     197  & 12  & $480$ & $2.7$ \\
$32^3\times 4$ & $4.30$  & $0.075$ & $0.001$ &     200  & 10  & $660$ & $3.8$ \\
$32^3\times 4$ & $4.50$  & $0.051$ & $0.001$ &     209  & 10  & $960$ & $5.5$ \\[1ex]
\hline\hline
\end{tabular}
\caption{
Ensembles and their parameters: We list the lattice size, the inverse gauge coupling $\beta_g$, the lattice constant $a$ in
fm, the statistics, the extent $L_5$ used for the domain wall fermions, the temperature $T$ in MeV and the ratio
$T/T_c$ (see \cite{Cossu:2015kfa,Tomiya:2016jwr} for details).}
\label{tab:ensembles}
\end{table}

As already discussed, we measure finite temperature spatial correlators in the
$z$-direction, as was first suggested in~\cite{DeTar:1987xb}.
To compare the results from our different ensembles we plot the correlators as
a function of the dimensionless combination 
\begin{align}
z\,T \; = \; (n_z a)/(N_t a) \; = \; n_z/N_t \; ,
\label{z_dimless}
\end{align}
where~$z$ is the physical distance in the correlators, $T$ the temperature,
$a$ the lattice constant, $n_z$ the distance in lattice units and $N_t$ the
temporal lattice extent. 

We project to zero-momentum by summing over all lattice sites in slices orthogonal
to the $z$-direction, i.e., we consider
\begin{equation}
C_\Gamma(n_z) = \sum\limits_{n_x, n_y, n_t}
\braket{\mathcal{O}_\Gamma(n_x,n_y,n_z,n_t)
\mathcal{O}_\Gamma(\mathbf{0},0)^\dagger}.
\label{eq:momentumprojection}
\end{equation} 
Obviously this is the lattice version of the continuum form in Eq.~(\ref{corrdef}).


\section{\label{sec:results}Results}

\begin{figure}
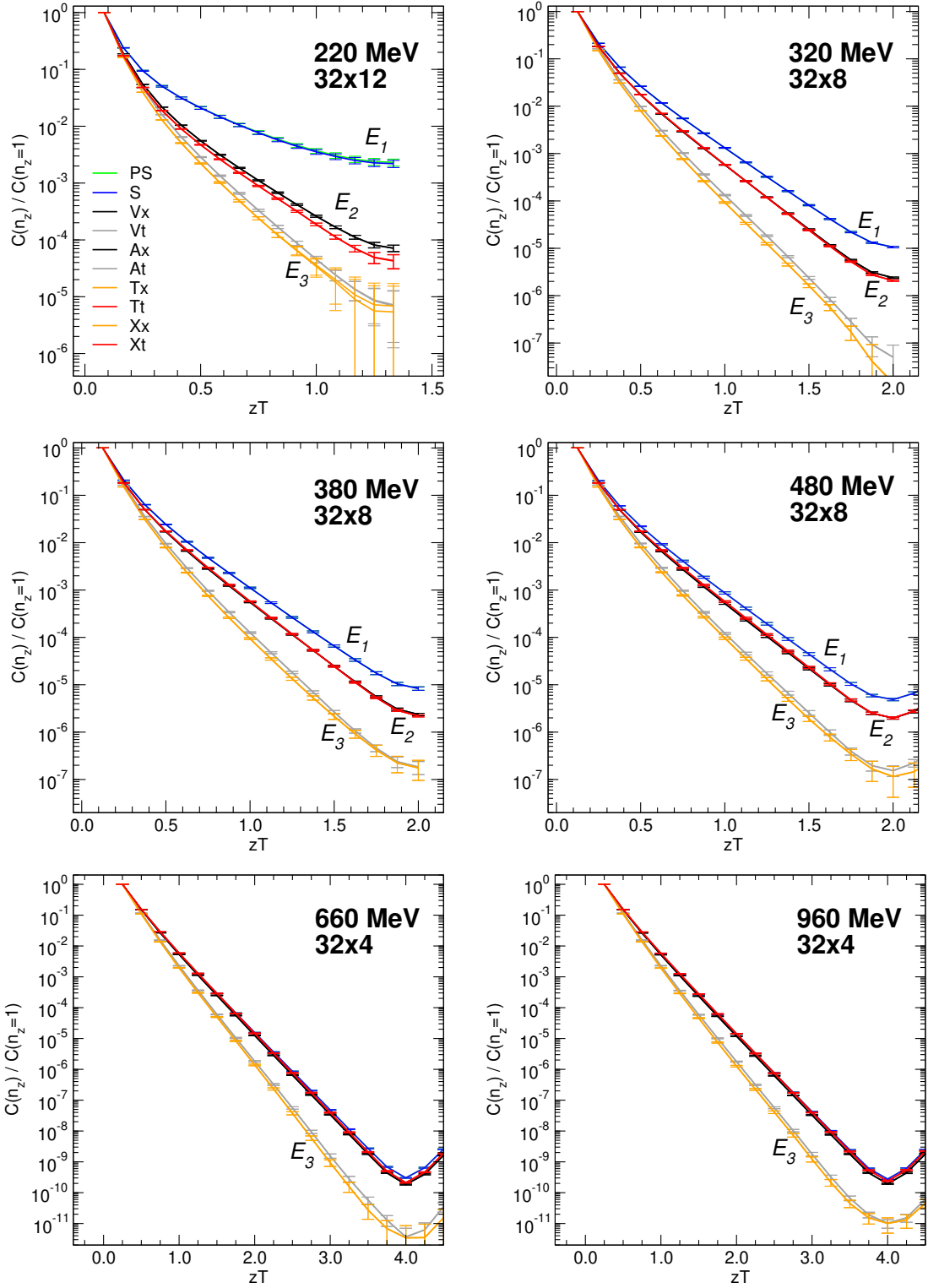

\centering
\hspace*{-10mm}
\includegraphics[scale=0.45]{{{1a_nt12}}} 
\hspace{4mm}
\includegraphics[scale=0.45]{{{1b}}}\\ 
\vskip3mm
\hspace*{-10mm}
\includegraphics[scale=0.45]{{{1c}}} 
\hspace{4mm}
\includegraphics[scale=0.45]{{{1d}}}\\ 
\vskip3mm
\hspace*{-10mm}
\includegraphics[scale=0.45]{{{1e}}} 
\hspace{4mm}
\includegraphics[scale=0.45]{{{1f}}}
\caption{
Overview of our spatial correlators in a wide range of temperatures. The correlators are shown as a function 
of the dimensionless combination $zT = n_z/N_t$ and are normalized to 1 at $n_z =1$. Note that the correlators 
are for different lattice sizes as indicated (compare Table~\ref{tab:ensembles} for details).
We label groups of correlators according to the multiplets $E_1, E_2$ and $E_3$ as introduced in 
Eqs.~(\ref{eq:e1}) -- (\ref{eq:e3}).}
\label{fig:allcorrs}
\end{figure}

In Fig.~\ref{fig:allcorrs} we compare the spatial correlators for a wide 
range of temperatures from $T \sim 220$ MeV to $960$ MeV to give an  impression of the changing behavior 
observed for different values of $T$.
The correlators are shown as a function of the dimensionless
combination $zT = n_z/N_t$ (compare Eq.~(\ref{z_dimless})) using the full range of $n_z$ values -- up to periodicity.
In order to compare different correlators without a proper renormalization, our correlators are normalized to 1 at 
$n_z = 1$. Because of the degeneracy of $x$ and $y$ components in vector operators
we show only the correlators for the $x$ components.
 
The top left panel of Fig.~\ref{fig:allcorrs} shows correlators at a
temperature of $T\sim 220$ MeV, i.e., $1.2\,T_c$.
All correlation functions of chiral partners are degenerate within errors.
In detail, this are the two pairs $(V_x,A_x)$ and $(V_t,A_t)$, each of which
reflects $SU(2)_R \times SU(2)_L$ symmetry.
$U(1)_A$ symmetry in the vector channel, represented by the operator pairs
$(T_x,X_x)$ and $(T_t,X_t)$, is manifest for all ensembles.
For the scalar $(PS,S)$ pair we find the restoration of $U(1)_A$ symmetry
to be heavily dependent on the parameters.
As it is evident from the top left panel of Fig.~\ref{fig:allcorrs},
$PS$ and $S$ are degenerate within errors for our finest lattice.
On the coarser $32\times 8$ ensemble at 220 MeV we find a
visible difference of $PS$ and $S$ correlators consistent with previous
findings in literature, e.g. the data for staggered quarks presented
in Fig. 7 of Ref.~\cite{Cheng:2010fe}.\footnote{
For detailed studies of $\mathrm{U}(1)_A$ symmetry around $T_c$
see e.g.~\cite{Brandt:2016daq} or~\cite{Tomiya:2016jwr}.
The latter study uses the same simulation setup as the present work.
}

For temperatures between $T \sim 220$ -- $500$ MeV the 
correlators are grouped into three distinct multiplets\footnote{Note that in $E_2$ and $E_3$ we leave out the 
$y$ components which are exactly degenerate with the respective $x$ components explicitly listed in 
$E_2$ and $E_3$.}:
\begin{eqnarray}
E_1: & \qquad PS \leftrightarrow S \; , \label{eq:e1} \\
E_2: & \qquad V_x \leftrightarrow T_t \leftrightarrow X_t \leftrightarrow A_x \; , \label{eq:e2} \\
E_3: & \qquad V_t \leftrightarrow T_x \leftrightarrow X_x \leftrightarrow A_t \; . \label{eq:e3}
\end{eqnarray}
Possible splittings within each of these multiplets are obviously much smaller
than the distances between the multiplets. The multiplet structure reflects the symmetries as follows: 
The multiplet $E_1$ indicates the restoration of~$U(1)_A$ symmetry. Degeneracies within the multiplets
$E_2$ and $E_3$ reflect the larger symmetries $SU(2)_{CS}$ and $SU(4)$
as discussed in the previous section.

The formation of the multiplet $E_3$ is not necessarily a consequence of the $SU(2)_{CS}$ and $SU(4)$ 
symmetries as the same degeneracy of correlators is seen also for non-interacting quarks (\ref{Corr_multiplets}) and can be attributed to current conservation. Consequently from the observation
of the $E_3$ multiplet alone we could not claim the emergence of the $SU(2)_{CS}$ and $SU(4)$ symmetries. 
However, the $E_2$ degeneracy is not manifest in the free quark system 
(\ref{Corr_multiplets})
and indeed can be attributed to the 
emergent $SU(2)_{CS}$ and $SU(4)$ symmetries.

We speak of separate multiplets when the splittings within the
multiplets are much smaller than splittings between different multiplets.
All correlators connected by chiral $U(1)_A$ and $SU(2)_L \times SU(2)_R$
transformations are indistinguishable at all temperatures.
At temperatures above $T \sim 600$ MeV we observe that the distinct multiplet $E_2$, related to emergence of the $SU(2)_{CS}$ and $SU(4)$ symmetries, is washed out. 
The remaining $E_3$ multiplet structure
can be attributed to quasi-free quarks.

\begin{figure}
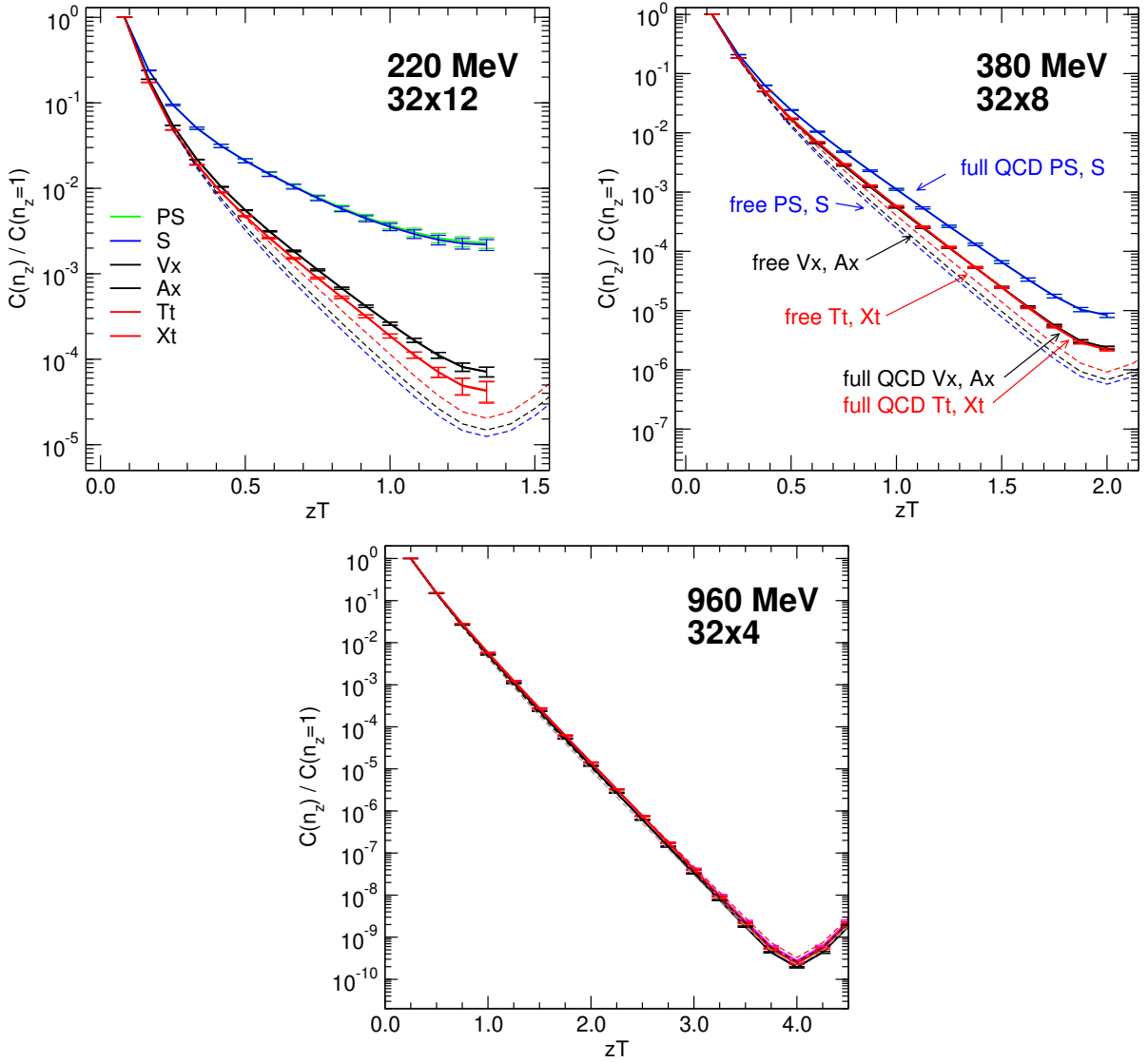

  \centering
  \includegraphics[scale=0.50]{{{2a_nt12}}} 
  \hspace{4mm}
  \includegraphics[scale=0.50]{{{2b}}} \\
  \vskip3mm
  \includegraphics[scale=0.50]{{{2c}}}
\caption{
Correlation functions of the bilinears in the $E_1$ and $E_2$ multiplets. The structure of the plots is the same
as described in the caption of Fig.~\ref{fig:allcorrs}, with the addition of the correlators for free quarks shown as 
dashed lines.}
\label{fig:e2_withfreedata}
\end{figure}

In Fig.~\ref{fig:e2_withfreedata} we now focus on the $E_1$ and $E_2$ multiplets at three different temperatures. 
For comparison we also show the corresponding correlators computed for free quarks (dashed lines). 
The latter correlators are obtained with the same lattice Dirac operator and lattice size as used for the full 
QCD but now with a unit gauge configuration. We note that for
free quarks only those degeneracies exist that are predicted by the chiral $U(1)_A$ and 
$SU(2)_L \times SU(2)_R$ symmetries. 
 
For the lowest temperature $T \sim 220$ MeV we still observe a small residual
splitting within the $E_2$ multiplet,
while at $T \sim 380$ MeV the difference nearly vanishes.
Furthermore, there is a clear splitting between the $E_1$ and $E_2$ multiplets
indicating  $SU(2)_{CS}$ and $SU(4)$ symmetries.
In addition all correlators are well separated from their free quark
counterparts shown as dashed curves.  

At the highest temperature of this study, $T \sim 960$ MeV, the situation has changed considerably:
All correlators almost perfectly coincide with the corresponding free
correlators, as seen by the dashed lines on top of the data points for the full QCD correlators.
Thus at $T \sim 960$ MeV we have reached the region where only chiral
$U(1)_A$ and $SU(2)_L \times SU(2)_R$ symmetries exist and the coincidence
with the free correlators suggests a gas of quasi-free quarks.

\begin{figure}
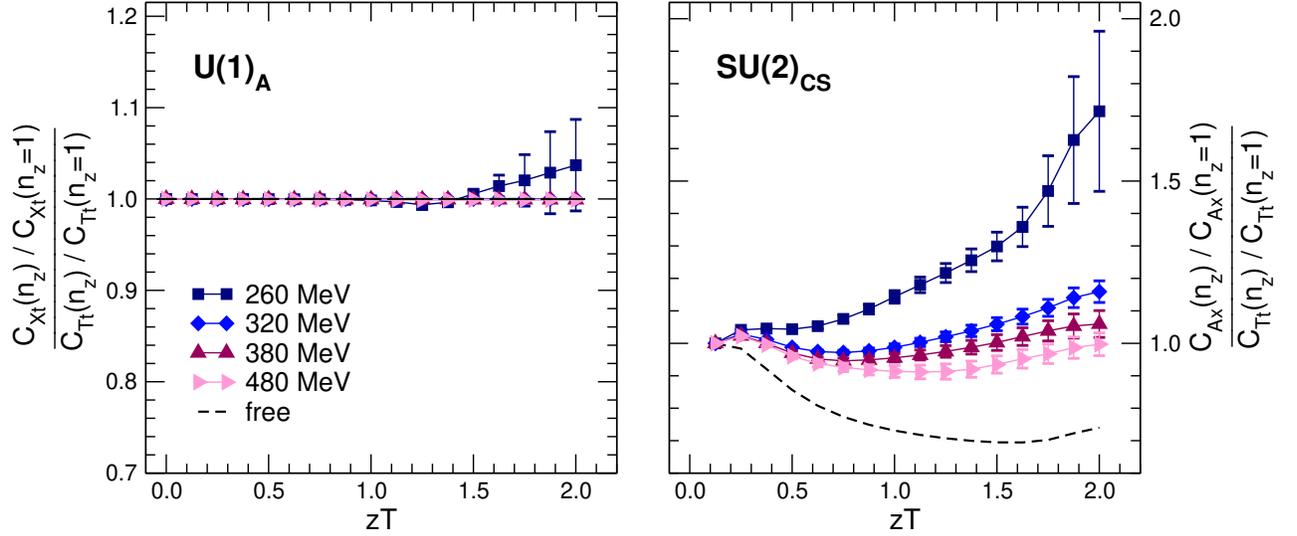

\centering
\includegraphics[scale=0.58]{{{3a}}}
\hspace{4mm}
\includegraphics[scale=0.58]{{{3b}}}
\caption{
Ratios of normalized correlators for different bilinears from the $E_2$ multiplet at different temperatures 
($32^3 \times 8$ lattices): The lhs.\ plot shows the ratio $C_{X_t}/C_{T_t}$, i.e., a ratio of correlators
connected by $U(1)_A$. The rhs.\ plot shows the ratio $C_{A_x}/C_{T_t}$, 
i.e., two correlators connected by $SU(2)_{CS}$ transformations. In both cases we show the corresponding 
ratios for free quarks as dashed curves.}
\label{fig:ratios_u1_su2cs}
\end{figure}

\begin{figure}
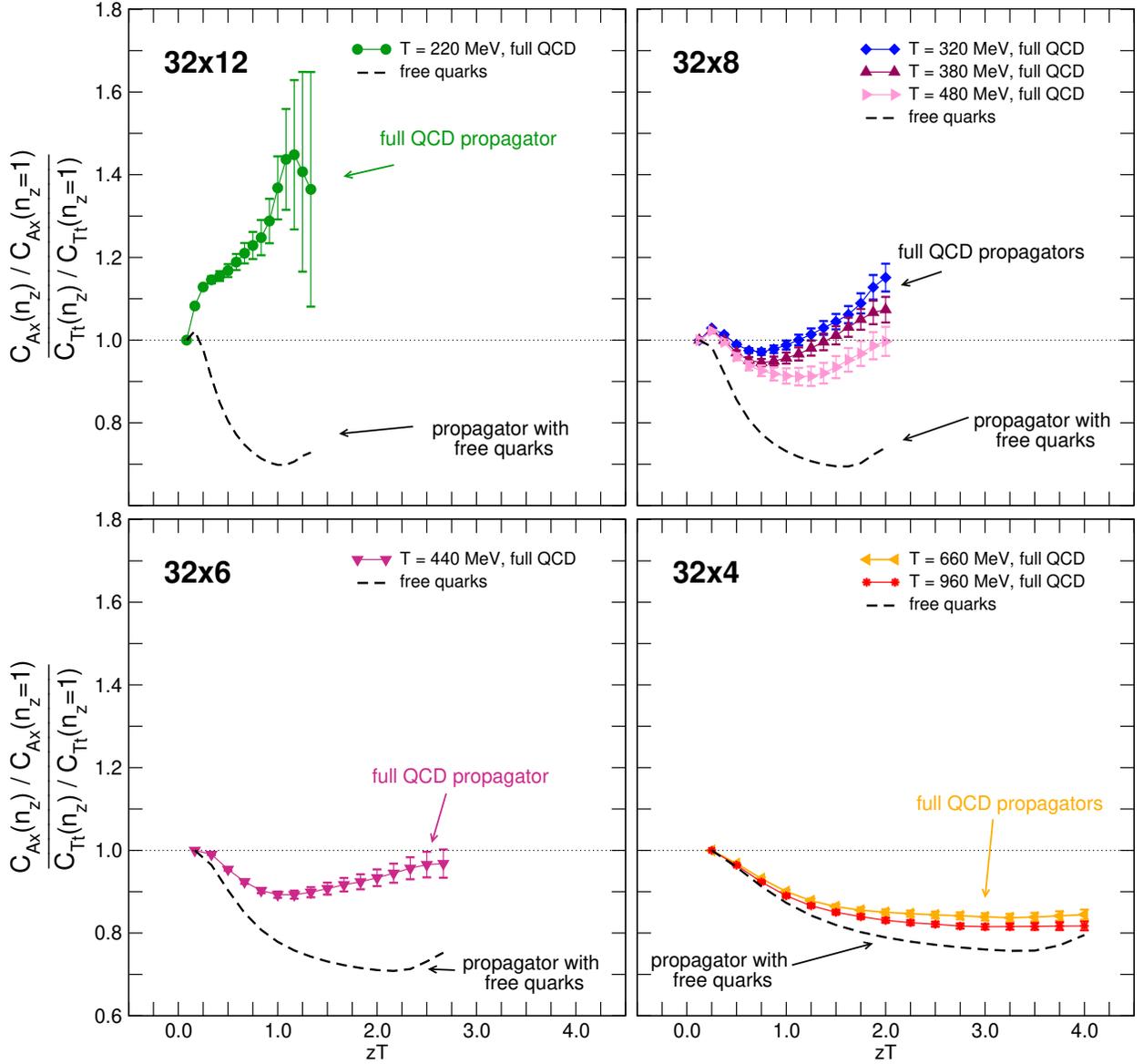

\centering
\hspace*{-9mm}
\includegraphics[scale=0.48]{{{4a}}} 
\includegraphics[scale=0.48]{{{4b}}} \\
\hspace*{-9mm}
\includegraphics[scale=0.48]{{{4c}}} 
\includegraphics[scale=0.48]{{{4d}}}
\caption{
The ratio $C_{A_x}/C_{T_t}$ for different temperatures. The two correlators from the $E_2$ multiplet are related by 
$SU(2)_{CS}$ and deviations from 1 indicate violation of the symmetry. The 4 different plots group together the results for
lattices with the same aspect ratio.}
\label{fig:ratios_all_su2cs}
\end{figure}

In an attempt of discussing the observed evolution of symmetries more
quantitatively, in Figures \ref{fig:ratios_u1_su2cs} and
\ref{fig:ratios_all_su2cs} we study ratios of correlators, where the fully
symmetric case corresponds to a constant ratio 1 for all $z$. 
In Fig. \ref{fig:ratios_u1_su2cs} we show ratios of normalized correlators for
different bilinears from the $E_2$ multiplet. 
The ratios are plotted as function of the dimensionless quantity
$zT = n_z/N_t$ and we compare different temperatures.

In the lhs.\ plot we show the ratio $C_{X_t}/C_{T_t}$.
The two correlators are related by $U(1)_A$ and a deviation from a constant
ratio~1 indicates a violation of $U(1)_A$.
The data shows no breaking effects within errors.

In the rhs.\ plot we show the ratio $C_{A_x}/C_{T_t}$.
These two correlators are related by $SU(2)_{CS}$ and thus a deviation
from~1 indicates a violation of exact $SU(2)_{CS}$. 
Here the lowest temperature displays sizable residual violation,
which gradually becomes smaller with increasing temperature.
At $T \sim 380$ MeV the deviation from 1 becomes minimal.

Finally, in Fig.~\ref{fig:ratios_all_su2cs} we analyze the $SU(2)_{CS}$ sensitive ratio $C_{A_x}/C_{T_t}$
for all our ensembles in a wider range of temperatures.
We observe an evolution from sizable deviation from 1
at the lowest temperature $T \sim 220$ MeV towards a coincidence with the corresponding ratio of correlators 
for free quarks at the highest temperature, i.e.~$T \sim 960$ MeV.
For intermediate temperatures  
we observe small deviations from 1.

Figs. \ref{fig:ratios_u1_su2cs} and \ref{fig:ratios_all_su2cs} demonstrate that
-- while the chiral symmetries are practically exact -- the $SU(2)_{CS}$ symmetry is not exact. 
Let us introduce a measure for the symmetry  breaking and find  a temperature range where the
symmetry is appropriate.

In general a symmetry is established via its multiplet structure.
For any multiplet structure a crucial parameter is the ratio of the
splitting within a multiplet to the distance between multiplets.
The splitting within a multiplet by itself
is irrelevant without a scale, and should be compared
to a scale relevant for the given problem,
e.g. the distance between multiplets.
Consequently, in our case the 
breaking of $SU(2)_{CS}$ and $SU(4)$ can be identified through the
parameter 
\begin{equation}\label{def_kappa}
  \kappa = \frac{|C_{A_x} - C_{T_t}|}{|C_{A_x} - C_{S}|}.
\end{equation} 
If $\kappa \ll 1$, then we can declare
 an approximate or -- if zero -- an exact
symmetry.
If $\kappa \sim 1$, the symmetry is absent.
The criterion of small $\kappa$ corresponds to the existence of a distinct
multiplet $E_2$ that should be well separated from the multiplet $E_1$.
From the free quark expression (\ref{Corr_multiplets}) one finds  $\kappa \sim 1$,
which stresses again that there is no chiral-spin
symmetry for free quarks. 

\begin{figure}
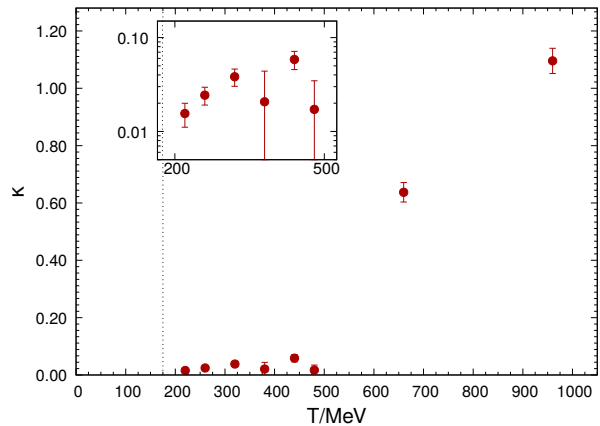

  \centering
  \includegraphics[scale=0.65]{{{kappa}}}
  \caption{The symmetry breaking parameter $\kappa$ defined in (\ref{def_kappa}) evaluated at $zT=2$ for different temperatures.
           The dashed line represents $T_c$.}
  \label{fig:kappa_zT2}
\end{figure}

In Fig. \ref{fig:kappa_zT2} we show the evolution of the symmetry breaking parameter $\kappa$ as
a function of temperature at $zT=2$.
The value of $\kappa$ is less than 5 \% for all ensembles with $T \sim 220$ -- $500$ MeV.
This implies that the symmetries that we observe in the range between
$T \sim 220$ MeV and   $500$ MeV are well  pronounced.

At temperatures between $T \sim 500$ MeV and  $T \sim 660$ MeV we notice a drastic increase
of the symmetry breaking parameter $\kappa$ to values of the order $\sim 1$.
We conclude that QCD exhibits an approximate $SU(2)_{CS}$ symmetry in the
temperature range between $T \sim 220$ -- $500$ MeV
($1.2T_c$ -- $2.8T_c$)
with symmetry breaking
less than 5\% as measured with $\kappa$.
This suggests that the $SU(2)_{CS}$ symmetric regime begins just after
the $SU(2)_R \times SU(2)_L$ restoration crossover.

We stress once more that the $SU(2)_{CS}$ symmetry is related to
different components of the strong interaction.
As we have discussed, an exact $SU(2)_{CS}$ symmetry implies that
the interaction is strictly chromo-electric. 
Thus the observed evolution of the $SU(2)_{CS}$ symmetry as a
function of temperature suggests the following picture for the
relevant degrees of freedom in high temperature QCD:
At  $T \sim 220$ MeV we find $C_{A_x}/C_{T_t} > 1$ and a small  violation
of $SU(2)_{CS}$ such that the interaction between the quarks must
be mediated not only by the chromo-electric component, but also to
some extent by the chromo-magnetic components of the gluonic field. 
When increasing the temperature, the ratio $C_{A_x}/C_{T_t}$ 
evolves towards 1. 
This implies that at $T \sim 380$ MeV the
chromo-magnetic interaction has become washed out and quarks interact
via the chromo-electric field.
The remaining small breaking of $SU(2)_{CS}$ is due to the quark
kinetic term.
It suggests that in this regime the elementary objects are chirally
symmetric quarks confined by the chromo-electric field.
At even higher temperatures also the contribution of the
chromo-electric interaction decreases
and the system enters the region of quasi-free quarks, as
reflected by the fact that for our highest temperatures the ratio
$C_{A_x}/C_{T_t}$ approaches the corresponding curve for free quarks.

We stress that the emerging $SU(2)_{CS}$ and $SU(4)$ symmetries,
observed in the range of $T \sim 220$ MeV to $T \sim 500$ MeV,
are incompatible with the picture of free deconfined quarks. 

This view is also reflected in the exponential decay properties
-- i.e., the factors $\propto \exp(-c\,z)$ -- 
of the full QCD correlators. A system of two free quarks cannot have $z$-correlators  where the 
exponent $c$ is smaller than twice the lowest Matsubara frequency $2\omega_0$, due to the anti-periodic
boundary conditions of fermions in time direction (compare Eq.~(\ref{Corr_multiplets_asympt})).  
If the exponent $c$ is smaller for the interacting case, this suggests that the 
quark-antiquark system is still coupled into a bosonic compound, since periodic boundary
conditions for bosons do allow for the exponent $c$ to be smaller than
$2\,\omega_0$.
Fig.~\ref{fig:e2_withfreedata} shows that the full $PS$- and $S$-correlators have significantly
smaller exponents $c$ than their non-interacting counterparts, which suggests that these
correlators correspond to coupled quark-antiquark compounds \cite{DeTar:1987xb}.
In the $J=1$ channels the difference of the exponents $c$ for full and free correlators at
temperatures $T < 500$ MeV is much smaller, but still visible, and suggests a residual binding also
in this case.


\section{\label{sec:conclusions}Conclusions}

In this paper we have studied spatial correlators of all
possible local $J=0$ and $J=1$ bilinears in high temperature lattice QCD.  We use $N_F=2$ flavors of domain
wall fermions and study temperatures up to $T \sim 960$ MeV.
Above the chiral restoration crossover at a pseudo-critical temperature
$T_c\sim 175$ MeV we observe restoration of chiral $SU(2)_L \times SU(2)_R$ symmetry
for all studied temperatures.
While $U(1)_A$ symmetry is present in all ensembles above 260 MeV, its
restoration at 220 MeV is observed on the finest lattice solely.

In the range between $T \sim 220$ MeV  and $T \sim 500$ MeV we observe the formation of multiplets
in spatial correlators that indicate larger emergent symmetries described by the chiral spin 
$SU(2)_{CS}$ and $SU(4)$ groups with the breaking effects below 5 \%
as measured by $\kappa$.  
These symmetries include the chiral $U(1)_A$ and $SU(2)_L \times SU(2)_R$ groups as well as transformations that 
mix the right- and left-handed components of quarks as subgroups. These are not symmetries of the free Dirac action
but are symmetries of the fermionic charge. In a given reference frame, which in our case is the medium rest frame,
the quark - chromo-electric interaction is invariant under both $SU(2)_{CS}$ and $SU(4)$ transformations,
while the quark - chromo-magnetic interaction as well as the quark kinetic term break them. 

The emergence of these symmetries in the $T\sim 220$ -- $500$ MeV window
($1.2T_c$ -- $2.8T_c$)
suggests that the chromo-magnetic interaction between quarks
is  screened
at these temperatures, while the confining chromo-electric
interaction is still active.
The emergence of approximate $SU(2)_{CS}$ and $SU(4)$ symmetries in the
window $T\sim 220$ -- $500$ MeV is the principal result of our study.
These emergent symmetries are incompatible
with the picture of free, deconfined quarks and suggest that the
physical degrees of freedom are chirally symmetric quarks bound by the
chromo-electric interaction without chromo-magnetic effects.
The latter conclusion is based entirely on our
lattice observations and the symmetry classification of the QCD Lagrangian,
i.e., it is model independent. We remark that correlation functions
with the $SU(2)_{CS}$ and $SU(4)$ symmetries cannot be analyzed perturbatively 
because perturbation theory reflects the symmetries of the
free Dirac equation. 

While we do not advocate any microscopic description of these ultrarelativistic objects,
they are reminiscent of ``strings". A string is the only known
mathematical description of purely electric, relativistic objects, though
a consistent theory of a relativistic string with  quarks at
the ends is missing in four dimensions.
We refer the $SU(2)_{CS}$ and $SU(4)$ symmetric regime at temperatures
$T\sim 220$ -- $500$ as the ``stringy fluid" to  emphasize the possible  nature of the
objects - chirally symmetric quarks bound by the  electric field.

At temperatures above $T \sim 600$ MeV these symmetries  disappear
and the QCD correlation functions approach  the correlators
calculated with free, non-interacting quarks. This suggests that only
at temperatures $T \sim 1$ GeV and above hot QCD matter can be 
approximately described as a gas of weakly interacting quarks
and gluons -- the Quark-Gluon Plasma (QGP).

Our analysis of spatial correlators and their multiplet structure suggests the following three
regimes of QCD when increasing the temperature: At low temperatures up to
the pseudo-critical temperature $T_c$ QCD matter is a hadron gas where all chiral
symmetries are broken by the non-zero quark condensate. 
From the hadron gas regime below $T_c$  
there is a crossover to a regime  
with approximate $SU(2)_{CS}$ chiral spin symmetry, where quarks are predominantly bound by
the chromo-electric interaction. This crossover coincides or is close
to the chiral $SU(2)_L \times SU(2)_R$ restoration crossover
(while in our setup the chiral crossover is at $T_c \sim  175$ MeV, for
three-flavor QCD the chiral crossover is at a somewhat lower temperature of
155 MeV \cite{Bazavov:2018mes}).
In the range $T \sim 500$ -- $660$ MeV ($2.8T_c$ -- $3.8T_c$) there is a fast increase
of symmetry breaking:
the confining electric interaction becomes small relative to the quark kinetic term.
Finally, up to $T \sim 1$ GeV ($5.7 T_c$) there is an evolution to a weakly interacting QGP, where the relevant 
symmetries are the full set of chiral symmetries. Fig.~6 provides an illustrative sketch of this temperature
evolution for the effective degrees of freedom of QCD.
We note that the temperature range, in which the most drastic changes of thermodynamical bulk quantities
occur, coincides qualitatively with the ``stringy fluid'' regime, see, e.g., Fig.~4 of Ref.~\cite{Bazavov:2017dsy}.
\begin{figure}
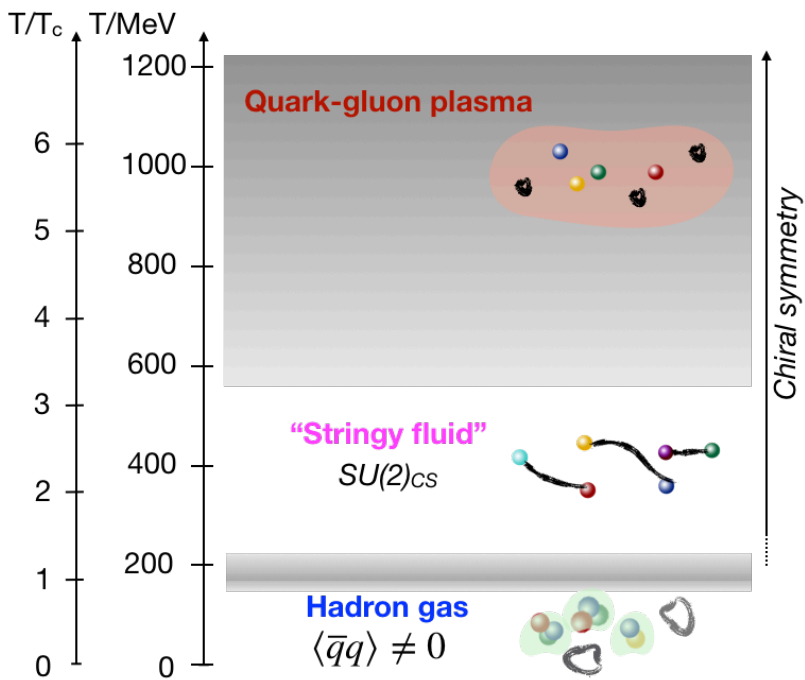

\centering
\includegraphics[scale=0.4]{{{fig_v}}} 
\caption{
Illustrative sketch for the temperature evolution of the QCD effective degrees of freedom as suggested by the changing 
symmetry content manifest in our spatial correlators.}
\label{fig:sketch}
\end{figure}


\begin{acknowledgments}
Support from the Austrian Science Fund (FWF) through the grants
DK W1203-N16 and P26627-N27, as well as from NAWI Graz is acknowledged.
The numerical calculations were performed on the Blue Gene/Q at KEK under its 
Large Scale Simulation Program (No. 16/17-14), at the Vienna Scientific Cluster (VSC)
and at the HPC cluster of the University of Graz. This work is supported in 
part by JSPS KAKENHI Grant Number JP26247043 and by the Post-K 
supercomputer project through the Joint Institute for Computational 
Fundamental Science (JICFuS).
S.P. acknowledges support from ARRS (J1-8137, P1-0035) and DFG (SFB/TRR 55).
\end{acknowledgments}


\section*{Appendix A}

All free spatial continuum correlators that we discuss in Section 2 can be expressed as linear combinations of  
$C_z(z)$ and $C_\tau(z)$ defined in Eq.~(\ref{CsCzCtau}). These two correlators can be simplified by 
switching to polar coordinates $p_x = r \cos(\varphi)$, $p_y = r \sin(\varphi)$. The $\varphi$-integration gives 
a factor of $2\pi$ and the transformation $\xi^2 = (r/\omega_n)^2 + 1$ of the remaining integration variable 
brings the correlators to the form 
\begin{eqnarray}
C_z(z) & = & \frac{1}{2\pi \beta} \sum_{n \in \mathds{Z}}  \omega_n^2 \int_1^\infty \!\! d\xi \, \xi \,  
e^{\, - 2 \, z \, |\omega_n| \, \xi}  \; ,
\nonumber \\
C_\tau(z) & = & \frac{1}{2\pi \beta} \sum_{n \in \mathds{Z}}  \omega_n^2 \int_1^\infty \!\! d\xi \, \xi \, \frac{1}{\xi^2} \, 
e^{\, - 2 \, z \, |\omega_n| \, \xi}  \; .
\label{sumintegral}
\end{eqnarray}
Both contain the Matsubara sum 
\begin{eqnarray}
\hspace*{-6mm} \sum_{n \in \mathds{Z}}  \omega_n^{\,2} \; e^{\, - 2 \, z \,  |\omega_n| \, \xi} & = & 
2 \! \sum_{n \in \mathds{N}_0} \! \omega_n^{\,2} \; e^{\, - 2 \, z \, |\omega_n| \, \xi} \; = \; 
\frac{1}{2 z^2} \frac{d^2}{d\xi^2}
\sum_{n \in \mathds{N}_0} e^{\, - 2 \, z \, |\omega_n| \, \xi} 
\label{matsubarasum}
\\
\hspace*{-6mm}&=& \frac{1}{2 z^2} \frac{d^2}{d\xi^2} e^{\, - z \, \frac{2\pi}{\beta} \, \xi} \!\! \sum_{n \in \mathds{N}_0} 
\!\!\! \left( \!e^{- 2 z \frac{2\pi}{\beta} \xi}\!\right)^{\!n} = \frac{1}{2 z^2} \frac{d^2}{d\xi^2} 
\frac{ e^{\, - z \, \frac{2\pi} {\beta}  \, \xi} }{ 1\!- \!e^{- 2 z \frac{2\pi}{\beta} \xi} } = \frac{1}{4 z^2} \frac{d^2}{d\xi^2}  
\frac{1}{\sinh(z \, \frac{2\pi}{\beta} \, \xi)} \; , 
\nonumber
\end{eqnarray}
where in the second step we have split the sum over $n \in \mathds{Z}$ into a positive and a negative part which can be 
transformed into each other by flipping the sign of $n$ and a trivial shift. Subsequently we generated the factor
$\omega_n^{\,2}$ with a second derivative and finally used the geometric series formula for the sum.  Below we will 
use both, the final expression as a derivative, as well as the other form of a sum over $n \in \mathds{N}_0$.

For solving the first integral $C_z(z)$ we use the form of the Matsubara sum (\ref{matsubarasum}) as a second 
derivative and insert this in (\ref{sumintegral}). Subsequently two partial integrations can be used to solve $C_z(z)$ 
in closed form ($\omega_0 = \pi/\beta$),
\begin{equation}
C_z(z) \, = \, \frac{\pi}{2 \beta^3} \, \frac{1}{2 \, z \omega_0} \, \frac{1}{\sinh(2 \, z \omega_0)} \left[
\mbox{cotanh}\, (2 \, z  \omega_0) + \frac{1}{2 \, z \omega_0} \right] \; .
\label{J1result}
\end{equation}
For the evaluation of $C_\tau(z)$ we keep the sum explicitly and find,
\begin{equation}
C_\tau(z) \; = \; \frac{1}{\pi \beta} \sum_{n \in \mathds{N}_0} \!\! \omega_n^{\,2} \! \int_{2 \, z  \omega_n}^\infty \!\!\!\! d\zeta \, 
\frac{e^{\, - \zeta}}{\zeta} \; = \;  \frac{1}{\pi \beta} \sum_{n \in \mathds{N}_0} \!\! \omega_n^{\,2} \, E_1(2  \, z  \omega_n) \; .
\label{J2result}
\end{equation}
In the first step we used the variable transformation $\zeta = 2\, z  \omega_n \xi$ which brings the integral 
into the standard form \cite{nist} for the exponential integral $E_1(x) \equiv \int_x^\infty d\zeta \, e^{-\zeta}/\zeta$.

We conclude this appendix with quoting the asymptotic forms for the integrals $C_z(z)$ and $C_\tau(z)$ which can be 
obtained with power series for standard functions from (\ref{J1result}) and the known expansion \cite{nist} for the 
exponential integral in (\ref{J2result}),
\begin{eqnarray}
C_z(z) & = & \frac{\pi}{\beta^3} \, \frac{e^{\, - \, 2 \, z  \omega_0}}{2 \,z \omega_0}
\left[ 1 + \frac{1}{2 \, z \omega_0} \right] \; + \; 
O \left( \frac{e^{\, - 4 \, z \omega_0}}{z\omega_0} \right) \; ,
\label{J_asymptotics} \\
C_\tau(z) & = & \frac{\pi}{\beta^3} \, \frac{e^{\, - \, 2 \, z \omega_0}}{2 \, z \omega_0}
\left[ 1 - \frac{1}{2 \, z  \omega_0} + O \left( \frac{1}{(z\omega_0)^2} \right)\right] \; + \; 
O \left( \frac{e^{\, - 6 \, z \omega_0}}{z \omega_0} \right) .
\nonumber
\end{eqnarray}


\end{document}